\documentstyle[12pt]{article}
\newcommand{\pn}{\par \noindent}
\newtheorem{theo}{\em Theorem}[section]
\newtheorem{definicja}[theo]{\em Definition}
\newtheorem{lemat}[theo]{\em Lemma}
\newtheorem{przykl}[theo]{\em Example}
\newtheorem{proposition}[theo]{\em Proposition}
\newtheorem{corollary}[theo]{\em Corollary}
\newenvironment{tw}{\begin{theo}\rm}{\end{theo}}
\newenvironment{df}{\begin{definicja}\rm}{\end{definicja}}

\newenvironment{prop}{\begin{proposition}\rm}{\end{proposition}}
\newenvironment{cor}{\begin{corollary}\rm}{\end{corollary}}

\begin{document}

\title{STRUCTURE OF MALICIOUS SINGULARITIES}
\author{Michael Heller\thanks{Correspondence address: ul.
Powsta\'nc\'ow Warszawy 13/94, 33-110 Tarn\'ow, Poland. E-mail:
mheller@wsd.tarnow.pl} \\
Vatican Observatory, V-00120 Vatican City State
\and Zdzis{\l}aw Odrzyg\'o\'zd\'z \and
Leszek Pysiak \and Wies{\l}aw Sasin \\
Department of Mathematics and Information Science, \\ Warsaw
University of 
Technology \\
Plac Politechniki 1, 00-661 Warsaw, Poland}
\date{\today}
\maketitle

\begin{abstract}
In this paper, we investigate relativistic spacetimes together
with their singular boundaries (including the strongest
singularities of the Big Bang type, called malicious
singularities) as noncommutative spaces. Such a space is defined
by a noncommutative algebra on the transformation groupoid
$\Gamma = \bar E \times G$, where $\bar E $ is the total space
of the frame bundle over spacetime with its singular boundary,
and $G$ is its structural group. We show that there exists the
bijective correspondence between unitary representations of the
groupoid $\Gamma $ and the systems of imprimitivity of the group
$G$. This allows us to apply the Mackey theorem to this case,
and deduce from it some information concerning singular fibres
of the groupoid $\Gamma $. At regular points the group
representation, which is a part of the corresponding system of
imprimitivity, does not have discrete components, whereas at the
malicious singularity such a group representation can be a single
representation (in particular, an irreducible one) or a direct
sum of such representations. A subgroup $K\subset G$, from which
--- according to the Mackey theorem --- the representation is
induced to the whole of $G$, can be regarded as measuring the
``richness'' of the singularity structure. In this sense, the
structure of malicious singularities is richer than those of
milder ones.

\end{abstract}

\section{Introduction}
Among various kinds of singularities one meets in studying
solutions to Einstein field equations there are some of
especially difficult character. They deserved names such as:
strong curvature singularities (Tipler, 1977a, b; Rudnicki et
al., 2002), crushing singularities (Eardley, 1979), malicious
singularities (Heller and Sasin, 2002). These classes do not
necessarily coincide, but there are singularities that belong to
all of them. Typical examples are the initial and final
singularities in Roberston-Walker-Friedman-Lema\^{\i}tre (RWFL)
cosmological models and the central singularity in the
Schawarzschild solution. They are usually described as elements
(ideal points) of various singular boundaries of spacetime, but
one knows hardly anything about their geometric nature besides
their defining properties, i.e., that the spacetime curvature
blows up as one approaches such a singularity, that it
compresses any suitably defined volume to zero, or that the
fibre in the frame bundle over the singularity degenerates to a
single point.  In the present paper, we make an attempt to say
something more about the malicious singularity by looking
directly into its geometric nature.

To do so we must leave the category of smooth manifolds within
which singularities can be reached only by a kind of a limiting
process. In this paper, we treat spacetime together with its
singular boundary as a noncommutative space.  Although such a
space is, in principle, nonlocal (the concepts of point and its
neighborhood are meaningless in it), it can be studied in terms
of representations of a certain noncommutative algebra in
Hilbert spaces. It turns out that such an approach can provide
some information about the nature of singularities. This method
has been elaborated in the framework of our program of studying
singularities (Heller and Sasin, 1994, 1996, 1999, 2002). The
present paper is a direct continuation of (Heller and Sasin
2002), we continue to focus on malicious singularities and, by
consequently applying to them the theory of representations, we
tray to obtain a deeper insight into their geometric and
physical nature.

The geometric context of our approach is the following. First,
we consider singularities as elements of the $b$-boundary of
spacetime (Schmidt, 1971). Although serious difficulties arise
when this construction is applied to spacetime with some strong
singularities, such as the ones in the closed RWFL model or in
the Schwarzschild solution (Bosshard, 1976; Johnson, 1977), they
can be overcome if the $b$-boundary construction is carried out
in the category of structured spaces (Heller and Sasin, 2002).
Schmidt's construction of the $b$-boundary of spacetime consists
in defining the Cauchy completion $\bar{E}$ of the total space
of the frame bundle $E$ over spacetime $M$ (with the help of a
Riemann metric on $E$), and ``projecting it down'' (by the
action of the Lorentz group that is a structured group of the
fame bundle) to obtain the $b$-completed spacetime $\bar{M} = M
\cup \partial_b M$, where $\partial_b M$ is the $b$-boundary of
spacetime $M$.  An element of $\partial_bM$ is said to be a {\it
malicious singularity\/} if the fiber in $\bar E$ over it
consists of a single point. Second, we construct the
transformation groupoid $\Gamma = \bar{E} \times G$, where $G$
is the structral group of the frame bundle, and a suitable
noncommutative algebra ${\cal A}$ on it. This algebra plays the
analogous role to the algebra of smooth functions on a manifold
and defines a noncommutative space that encodes ``noncommutative
properties'' of spacetime with singularities. It is a standard
method of changing the usual space into a noncommutative space
(Connes, 1994, pp. 99-103).  And finally, we use the groupoid
representation theory (or a representation of a suitable algebra
on this groupoid) to investigate the structure of both singular
and nonsingular groupoid fibres. Happily enough, there exists
the bijective correspondence between unitary representations of
the transformation groupoid $\Gamma =\bar E \times G$ and the
systems of imprimitivity of the Lie group $G$. This fact allows
us to reduce the problem to the study of unitary representations
of the group $G$ in a Hilbert space (which are better known than
groupoid representations).

Our main result is that at any regular (nonsingular) point of
spacetime the unitary representation, being a part of the system
of imprimitivity of the group $G$, does not contain discrete
components ($G$ has no discrete series of representations). This
is not true as far as malicious singularities are concerned. In
this case, the condition for the system of imprimitivity is
satisfied trivially. In particular, the corresponding group
representation can be a single irreducible representation or a
direct sum of such representations.

Spacetimes with $b$-boundaries are truly malicious geometric
objects. This is demonstrated, among others, by the fact that
the initial and final singularities in the closed RWFL
cosmological model form a single ``point'' in the corresponding
$b$-boundary, and are not Hausdorff separated from the rest of
spacetime (Bosshard, 1976; Johnson, 1977). Our analysis does not
change these conclusions, but the geometric tools used by us are
powerful enough to give us some insight into such seemingly
untractable situation. It is no longer a pathology, but rather a
mathematical structure that could be used, if necessary, to
model physical reality.

The organization of our material is the following. In section 2,
we present some elements of the groupoid structure. Our notation
is basically the same as in (Heller and Sasin, 2002), but to
make the present paper self-consistent we repeat some
definitions and prepare necessary tools from the theory of group
representations and systems of imprimitivity. In section 3, we
establish the bijective correspondence between representations
of the transformation groupoid $\Gamma = \bar E \times G$ and
the system of imprimitivity of the group $G$. Our main results,
concerning both regular points and malicious singularities, are
obtained in section 4, and are illustrated on a typical example
of the two-dimensional closed RWFL world model in section 5.
Finally, in section 6, we collect some comments and
interpretative remarks.

\section{Mathematical Preliminaries}
\subsection{Groupoids and Their Representations}
We begin with a brief description of the groupoid concept [see,
for instance, (Paterson, 1999, chapter 1)] mainly to fix
notation. {\it Groupoid\/} is a set $\Gamma $ with a
distinguished subset $\Gamma ^2 \subset \Gamma
\times \Gamma $, called the {\it set of composable elements\/},
together with two mappings:

$\cdot :\Gamma^2 \rightarrow \Gamma $ defined by $(x, y)
\mapsto x \cdot y, $ called {\it multiplication\/}, and
 
$^{-1}:\Gamma \rightarrow \Gamma $ defined by $x \mapsto
x^{-1}$ such that $(x^{-1})^{-1} = x$, called {\it inversion\/}.

Both mappings are supposed to satisfy the following conditions

(i) if $(x,y),(y,z) \in \Gamma^2$ then $(xy, z), (x, yz) \in
\Gamma^2 $ and $(xy)z = x(yz)$,

(ii) $(y,y^{-1}) \in \Gamma^2 $ for all $y \in \Gamma $, and if
$(x,y) \in \Gamma^2$ then $ x^{-1}(xy) = y$ and $(xy)y^{-1} =
x.$

We also define the {\it set of units\/} $\Gamma^0 = \{xx^{-1}: x
\in \Gamma \} \subset \Gamma , $ and the following
mappings: 

$d: \Gamma \rightarrow
\Gamma^0 $ by $d(x) = x^{-1}x,$ called {\it source mapping\/}, and 

$r: \Gamma \rightarrow \Gamma^0$ by $r(x) = xx^{-1}$, called {\it
target mapping}.  

Let us notice that  $(x,y) \in \Gamma^2$ if and only if $d(x)
= r(y)$.  

For each $u \in \Gamma^0$ we define the sets
$$\Gamma_u = \{x \in \Gamma: d(x) = u\} = d^{-1}(u) $$
and
$$\Gamma^u = \{x \in \Gamma: r(x) = u\} = r^{-1}(u).$$
Both these sets give different fibrations of $\Gamma $. The
set $\Gamma^u_u := \Gamma^u \cap \Gamma_u$ is closed under
multiplication and inverse. It is called the
{\it isotropy group\/} at $u$.

The above construction is purely algebraic, but it can be
equipped with the smoothness structure. In this case, it is
called a {\it smooth\/} or {\it Lie groupoid\/} (Paterson, 1999,
chapter 2.3).

The so-called transformation groupoids (or action groupoids)
form an important class of Lie groupoids.  Let $E$ be a
differential manifold with a group $G$ acting on it to the
right, $E \times G \rightarrow E$. This action leads to the
bundle $(E, \pi_M , M = E/G)$.  The Cartesian product $ \Gamma =
E \times G$ has the structure of a groupoid, and is called a
{\em transformation groupoid\/}. The elements of $\Gamma $ are
pairs $\gamma =(p,g)$ where $p\in E$ and $ g\in G$. Two such
pairs $\gamma_1 = (p, g)$ and $\gamma_2 =(pg, h)$ are composed
in the following way 
$$
\gamma_2 \gamma_1 = (pg, h)(p, g) = (p, gh).
$$
The inverse of $(p,g)$ is $(pg, g^{-1}).$ We could think on
$\gamma =(p,g)$ as on an arrow beginning at $ p$ and ending at
$pg$. Two arrows $\gamma_1$ and $\gamma_2$ can be composed
if the beginning of $\gamma_2$ coincides with the end of
$\gamma_1.$ 

The set of units is
$$
\Gamma^0 = \{\gamma^{-1} \gamma: \gamma \in \Gamma \} = \{(p,
e):\, p \in E \}.
$$

We shall often consider the ``fibres'' of this groupoid
$$
\Gamma_{(p,e)} = \{(p,g):g \in G\},
$$ $$
\Gamma^{(p,e)} = \{(ph^{-1},h): h \in G\}.
$$
In the following, we shall abbreviate the symbols
$\Gamma_{(p,e)}$ and $\Gamma^{(p,e)}$ to $\Gamma_p$ and
$\Gamma^p$, respectively.  If an element $\gamma = (p,g) \in
\Gamma $ is represented as an arrow from $p$ to $pg$, the set
$\Gamma_{p}$ should be thought of as the set of arrows which
begin in $(p, e)$, and the set $\Gamma^{p,}$ as the set of
arrows which end at $(p,e)$.

In what follows, we shall assume that $G$ is a unimodular group
(the Haar measure exists on $G$). Since all fibres of the
groupoid $\Gamma $ are isomorphic with $G$, the Haar system can
be defined on $\Gamma $, and $\Gamma $ can be regarded as a
locally compact Hausdorff groupoid (Paterson, 1999, p.  32).

Let us now recall the definition of a groupoid representation.
Let $\Gamma $ be a locally compact groupoid, $\Gamma^0$ its
space of units, and $(\Gamma^0,\{H_u\}_{u\in \Gamma^0},\mu )$ a
Hilbert bundle.  Here $\{H_u\}$ is a collection of Hilbert
spaces with $ u$ ranging over $\Gamma^0$, and $\mu $ is a
probability measure on $\Gamma^0$. By a {\em section\/} of the
Hilbert bundle we mean a function $f:\Gamma^0 \rightarrow
\bigcup_{u\in\Gamma^0}H_u$ where $f(u)\in H_u$.

\begin{df}
A {\em representation\/} ${\cal U}$ of the locally compact
groupoid $\Gamma$ is given by a Hilbert bundle
$(\Gamma^0,\{H_u\}_{u\in \Gamma^0},\mu )$, where $\mu$ is a
quasi-invariant measure on $\Gamma^0$, and a mapping $\Gamma
\ni x\rightarrow L(x)\in B(H_{d(x)},H_{r(x)})$, where $d$ and $r$ are
the source and the range mappings, respectively.  $L$ is
supposed to satisfy the following conditions

(i) $L(u)={\rm id}_{H_u},\,u\in\Gamma^0$,

(ii) $L(x)L(y)=L(x\cdot y)$, almost everywhere with respect of
the groupoid measure, for all $x,y\in\Gamma$ that can be
composed with each other (in the case considered in the present
paper this condition is satisfied everywhere),

(iii) $L(x)^{-1}=L(x^{-1})$, almost everywhere, for every $ x\in
G$,

(iv) for any two sections $\xi ,\eta\in
(L^2(\Gamma^0,\{H_u\}_{u\in \Gamma^0},\mu )$ of the Hilbert
bundle, the function
\[x\rightarrow (L(x)\xi (d(x)),\eta (r(x)))\]
is measurable on $\Gamma$ (Landsman, 1998, pp. 282-285).
\end{df}

\subsection{Induced Representations and Systems of Imprimitivity}

Let $G$ be a unimodular Lie group and $K$ its closed subgroup
(therefore, $K$ is also unimodular). In such a case, there
exists on $M = K\backslash G$ (the set of right cosets) a
$G$-invariant measure. Let further $(L,V)$ be a unitary
representation of the group $K$ in a Hilbert space $V$ (which
can also be finitely dimensional). Now, we form the Hilbert
space $H_L = L^2(M, V, d\mu )$ of a new representation of the
group $G$. $H_L$ consists of functions defined on $G$ with
values in $V$ such that

(i) the function $g \rightarrow (f(g), v)$, for every $g \in G$
and $v \in V$, is measurable (with respect to the Haar measure
$dg$ on $G$),

(ii) $f(kg) = L(k)f(g)$ for every $k\in K$ and $g \in G$
(covariance condition),

(iii) $\int_M ||f(g)||^2 d[g] < \infty $ where $[g] = Kg$.

The space $H_L$ with thy scalar product
\[ (f|f^{\prime })_{H_L} = \int_M (f(g),f^{\prime }(g))_V d[g]
\] 
is indeed the Hilbert space.

Let us define the operator
\[ U^L(g_0)f(g) = f(gg_0) . \]

\begin{df}
The representation $(U^L, H_L)$ of
the group $G$ is called the {\it induced representation\/} of
$G$ from the subgroup $K$ through the representation $(L,V)$.
\end{df}

This representation is unitary with respect to the above scalar
product (this follows from the invariance of the measure). Let
us notice that the regular representation\footnote{Let us recall
that the {\it left regular representation\/} of a unimodular Lie
group $G$ in the Hilbert space $L^2(G,dg)$ is given by
$U(g)=L_g$ where $L_gf(x):=f(g^{-1}x)$, and the {\it right
regular representation\/} of $G$ by $U(g)=R_g$ where
$U(g)f(x)=f(xg)$, for every $g\in G$.} $(R, L^2(G))$ of the
group $G$ is the induced representation from the trivial
subgroup $\{e\}$ by $(L,V)$ with $L=1$ and $V={\bf C}$.

Let again $G$ be a unimodular Lie group, $(U,H )$ its unitary
representation in a Hilbert space $H$, and $M$ a $G$-space,
i.e., a space with a (right) action of $ G$ (the action is not
necessarily transitive). Let further $P$ be a spectral measure
on $M$, i.e., a measure on Borel subsets of $ M$ with values in
the space of projection operators in the Hilbert space $ H$. If
$B\subset M$ is a Borel subset then $P(B)$ is an orthogonal
projection in $ H$.

\begin{df}
A quadruple $(G,U,M,P)$ is a {\em system of
imprimitivity\/} (S.I. for short) of the group $ G$ for the
representation $U$ with the base $M$ if the following conditions
are satisfied

(i) $P(M)={\rm id}_H,$

(ii) $U(g)P(B)U(g^{-1})=P(Bg^{-1})$

\noindent
for every $g\in G$ and $B\subset M$, $B$ being a Borel set.
\end{df}
Condition (ii) expresses a ``covariance'' of $P$ with respect to
$U$. S.I. is said to be {\em transitive\/} if $G$ acts
transitively on $ M$. In such a case, $M=K\backslash G$ where
$K$ is a closed subgroup of $G$.

There exists another (equivalent) definition of S.I. 

\begin{df}
S.I. of the group $G$ for the representation $ U$ with the base
$M$ is the quadruple $(G,U,M,\pi )$ where $(\pi ,H)$ is a
nondegenerate representation of the $*$-algebra $C_0(M )$ of
continuous functions on $M$ vanishing at infinity in a Hilbert
space $H$.  Conditions (i) and (ii) from the previous definition
are now replaced by
\[U(g)\pi (f)U(g^{-1})=\pi (R_gf)\]
where $R_gf(x)=f(xg)$, $x\in M,\,g\in G ,$$\,f\in C_0(M)$. S.I.
defined in this way is said to be {\em smooth\/} if $(\pi ,H)$
is a nondegenerate representation of the algebra $
C_0^{\infty}(M)$ of smooth functions on $M$ vanishing at
infinity.
\end{df}
\begin{tw}
{\em (Mackey).\/} If $(G,U,M,P)$ is a transitive S.I.
(i.e., $M=K\backslash G$) then the representation $(U,H)$ of the
group $G$ is induced from its subgroup $K$ or, more precisely
there exists a unitary representation $(L,U^L)$ of the subgroup
$K \subset G$ and the isomorphism of Hilbert spaces
$J:H\rightarrow H_L$ such that
\[JU(g)J^{-1}=U^L(g),\]
\[JP(B)J^{-1}=P^L(B)\]
for every $g\in G$ and every Borel subset $ B\subset M$. In
other words the representations $U$ and $U^L$ are unitary
equivalent (Mackey, 1952). $\Box$
\end{tw}
\section{Systems of Imprimitivity and Representations of the
Transformation Groupoid}
In this section, we find the correspondence between
representations of the transformation groupoid $\Gamma = E
\times G$ and systems of imprimitivity of the group $G$. It is
given by the following theorem.

\begin{tw}
Let $(G,U,X,\pi )$ be the S.I.
of the group $ G$ for the representation $U$ with base $X$, and
let ${\cal U}$ be a representation of the transformation
groupoid $\Gamma =X\times G$. There exists a one-to-one
correspondence
\[\{(G,U,X,\pi ))\leftrightarrow \{{\cal U}
\}.\]
\end{tw}
\pn
{\em Proof\/} is a combination of theorem 3.4.4, corollary 3.4.6
and corollary 3.7.4 from (Landsman, 1998), and theorem 3.1.1 from
(Paterson, 1999) [see also formula (3.20) from the book by Paterson].
$\Box$

We shall now directly construct the above correspondence for the
differential groupoid $\Gamma =\bar {E}\times G$.

{\em Step 1.\/} In this step we will construct another
realization of S.I. for our case. We choose a point $p_0 \in E$
such that $\tau (p_0)=m$, where $\tau : E \rightarrow M$ is the
canonical projection, and construct the space

\[{\cal F}_G(E_m,H)= \{\psi: E_m \rightarrow H:\psi (p_
0g)=U(g^{-1})\psi (p_0)\}\] 
where $H$ is a Hilbert space of the representation $ U$ (or of
$\pi$). The space ${\cal F}_G$ consists of continuous
functions.\footnote{Strong continuity is assumed as a part of
the definition of the unitary representation of a Lie group,
i.e., it is assumed that, for every $h\in H$, the function $G\in
g \rightarrow U(g)h\in H$ is continuous.} We equip this space
with the scalar product
\[ (\psi_1| \psi_2) = (\psi_1(p_0), \psi_2(p_0))_H \]
changing it into a Hilbert space.  We define the operator $\bar
U$ on the space ${\cal F}_G$
\[[\bar {U}(g)\psi ](p)=U(g)\psi (p),\]
and the representation $\bar {\pi }$ of $C_0(E)$ in the space
${\cal F}_G$
\[ [\bar{\pi }(f)\psi ](p_0g)=\pi (R_{g^{
-1}}f)\psi (p_0g)\] for every $\psi \in {\cal F}_G(E_m, H)$,
$f\in C_0(E)$, and for a point $p_0$ such that $ \tau (p_0)=m$;
that is to say
\[ [\bar {\pi }(f)\psi ](p_0)= \pi(f)\psi (p_0).\]
This condition enforces $(G, \bar U, E, \bar {\pi })$ to be an
S.I.

\begin{prop}
$(G,\bar {U},E,\bar{\pi })$ is an S.I. of the group $G$ for the
representation $\bar {U}$with base $E$.
\end{prop}
\pn
{\em Proof.\/} By using the covariance of the S.I. $(G,U,E,\pi
)$ and the properties of functions from ${\cal F}_G$ we check
the condition
\[\psi \in {\cal F}_G \Rightarrow \bar {\pi }(f)\in {\cal
F}_G \] and the covariance condition for
$(G,\bar{U},E,\bar{\pi })$.  $\Box $

{\it Step 2.\/} First, we construct a Hilbert spaces which will
form the Hilbert bundle. Let $p_0\in E$, and $ p_1=p_0g_0$.
\[{\cal H}^{p_0}=\{F:\Gamma^{p_0}\rightarrow 
H:\,F(p_0g^{-1},g)=U(g)F(p_0,e)\}.\]
Of course, functions $F$ are continuous, and we have the Hilbert
bundle $(E, {\cal H}^p, d\mu )$ where $d\mu $ is the measure on
$E$. 

Now, we define the representation operator of the groupoid $
\Gamma =E\times G$
\[{\cal U}(p_0,g_0):{\cal H}^{p_0}\rightarrow 
{\cal H}^{p_1}\]
by
\[[{\cal U}(p_0,g_0)F]=F(\gamma^{-1}\eta 
)=F(p_1g^{_{-1}},gg_0^{-1}).\]
Here $\gamma =(p_0,g_0)$, $\eta =(p_1g^{
-1},g)$.

Unitarity of the operator ${\cal U}(p_0, g_0)$ is implied by the
definitions of the scalar products in ${\cal H}^{p_0}$ and
${\cal H}^{ p_1}$
\[(F_1,F_2)_{{\cal H}^{p_0}}=(F_1(p_0,e)
,F_2(p_0,e))_H,\]
\[(\bar {F}_1,\bar {F}_2)_{{\cal H}^{p_1}}
=(\bar {F}_1(p_1,e),\bar {F}_2(p_1,e))_H.\]

We can easily check that all conditions of the groupoid
representation are satisfied (in this case, ``almost
everywhere'' is replaced by ``everywhere''). 

{\it Step 3.\/} Now, we should check that the constructed
groupoid representation corresponds to the initial S.I.
To this end, let us define the isomorphism of Hilbert spaces
\[J_{p_0}:{\cal H}^{p_0}\rightarrow {\cal F}_
G(E_m,H),\]
where $m=\tau (p_0)$,  by
\[\psi (p_0g^{_{-1}})=F(p_0g^{_{-1}},g)\]
where $J_{p_0}F=\psi$.

\begin{tw}
The isomorphisms $J_ p$ ``transform operators ${\cal U}(p_0,g_
0)$ onto operators $\bar {U}(g_0^{-1})$'' in the sense that $
{\cal U}(p_0,g_0)=J^{-1}_{p_1}\circ\bar {U}(g^{ -1})\circ
J_{p_0}$. In other words, the following diagram commutes

\vspace{1cm}
\unitlength=1mm
\special{em:linewidth 0.4pt}
\linethickness{0.4pt}
\begin{picture}(89.00,113.00)
\put(35.00,110.00){\makebox(0,0)[cc]{${\cal H}^{p_0}$}}
\put(35.00,80.00){\makebox(0,0)[cc]{${\cal H}^{p_1}$}}
\put(35.00,105.00){\vector(0,-1){19.00}}
\put(41.00,110.00){\vector(1,0){39.00}}
\put(85.00,110.00){\makebox(0,0)[lc]{${\cal F}_G(E_m,H)$}}
\put(41.00,80.00){\vector(1,0){39.00}}
\put(85.00,80.00){\makebox(0,0)[lc]{${\cal F}_G(E_m,H)$}}
\put(85.00,105.00){\vector(0,-1){19.00}}
\put(60.00,113.00){\makebox(0,0)[cb]{${\cal J}_{p_0}$}}
\put(60.00,77.00){\makebox(0,0)[ct]{${\cal J}_{p_1}$}}
\put(31.00,96.00){\makebox(0,0)[rc]{${\cal U}(p_0,g_o)$}}
\put(89.00,96.00){\makebox(0,0)[lc]{$\bar{U}(g_0^{-1})$}}
\end{picture}
\vspace{-6.5cm}
\end{tw}
\pn
{\it Proof\/} is by direct computation. $\Box $

\section{Systems of Imprimitivity for Sin\-gu\-lar Space\-times}
Let us notice that the groupoid $\Gamma$ is the disjoint sum of
$\Gamma_m=E_m\times G$, i.e., $\Gamma =\bigcup_{m\in
M}\Gamma_m$. And if the malicious singularity is present at $
m_1$, $\bar{\Gamma }=\bigcup_{m\in M}\Gamma_m \cup\Gamma_{
m_1}$ where $\Gamma_{m_1}=\{(0,0,\ldots 0)\}\times G$.

\begin{df}
Let $m_0\in\bar {M}$, and $\bar M = M \cup \{m_1\}$. The ${\em
local}$  S.I. at the point $m_0$ of the group $G = SO(3,1)$ for
the representation $(U,H)$ of $G$ is $(G,U, \bar{E}_{m_0}, \pi
)$.  Let us notice that the base of this S.I. is $E_{m_0}$.
\end{df}
\begin{prop}
Let $m\in M$ be a regular point. The S.I.
$(G,U,E, \pi )$ determines the local S.I. at the point $m$:
$(G,U,E_m, \pi_1 )$.
\end{prop}
\pn
{\it Proof.\/} Let us consider the algebra $C_0(E_m)$ of
continuous functions on $E_m$ vanishing at infinity. We choose a
point $p_0 \in E_m$, and want to show that $f\in C_0(E_m)$
can be ``extended'' to $\tilde f \in  C_0(E)$.

Let $\{({\cal O}_{n},f_{n})\}_{n\in {\bf N}}$ be the
approximate unit for the algebra $C_0(M)$; ${\cal O}_{n}$
is here a sequence of sets such that the closure $\bar{{\cal
O}}_n$ of each of them is compact, and supp$f_n \subset {\cal
O}_n$. We also assume that every ${\cal O}_n$ is the domain of
trivialization of the bundle $E \rightarrow M$. Let further, 
\[ \tilde {f_n}(m,g) = f_n(m)\cdot f(p_0g).\]

Of course, $\tilde {f}_n \in C_0(E)$. Finally, we define the
representation $\pi_1$ of the algebra $C_0(E_m)$ in the space
$H$
\[\pi_1(f) = \lim_{n\rightarrow \infty }\pi
(\tilde{f}_n) \]
where the limit is understood in the sense of strong
topology on the Hilbert space $H$.
$\Box $

\begin{tw}
Let $(G, \bar U, E_m, \bar {\pi })$ be a local
S.I. at a regular point $m\in M$. Then the representation $(\bar
U, {\cal F}_G(E_m, H))$, and consequently the representation
$(U,H)$, is unitary equivalent to the factor representation of
the regular representation of the group $G$ in the Hilbert space
$L^2(G)$.
\end{tw}
\pn
{\it Proof.\/} Let us notice that $E_m = K\backslash G$ where
$K= \{ e\}$. Therefore, the considered S.I. is transitive. On
the strength of the Mackey theorem, the representation $(U,H)$
is equivalent to the induced representation from the subgroup $K
= \{ e\}$. The inducing representation is given by the operator
$L={\rm id}_V$. (If the subgroup is trivial, the only
representation operator is the multiplication by 1, but the
representation space $V$ can be $n$-dimensional.) Consequently,
the representation induced by $L$ is given by the factor
representation containing the regular representation in the
space $L^2(G)$ with the multiplicity equal to dim$(V)$. $\Box $

\begin{cor}
The representation $(U,H)$, being a part of
the local S.I., does not contain discrete irreducible components.
\end{cor}
\pn
{\it Proof.\/} The regular representation of the group
$G=SO(3,1)$ has no discrete series. $\Box $

Let us notice, however, that this result depends on the
dimension of space. The group $SO(n,1)$ has no discrete series
for $n=2k+1$, but it has the discrete series for $n=2k$.

Let us now consider the situation in the malitiously singular
fiber; such a fiber is $\Gamma_{m_1} = \{ pt \} \times G$ where
$\tau(pt) = m_1 \in \bar M \setminus M$. In fact, $\Gamma_{m_1}$
can be regarded as a well defined groupoid
(indeed, $(pt, g_1)\circ (pt, g_2) = (pt, g_2g_1)$), and we can
consider the space ${\cal H}^{pt}$. If $F \in {\cal H}^{pt}$ then
\[F(pt,g)=U(g)F(pt,e).\]

We see that the operator $U(g)$ acts according to the rule, but
in the trivial way.  The same is true for the operator of the groupoid
representation
\[ [{\cal U}(pt,g_0)F](pt,g) = F(pt, gg_0^{-1}) =
U(g_0^{-1})F(pt,g).\] 

Let now $(G,U, E_{m_1}, \pi )$ be the local S.I. at the point
$pt$. We have $C_0(E_{m_1}) \simeq {\bf R}$, and the condition of
imprimitivity
\[ U(g)\pi (f) U(g^{-1}) = \pi (f), \, f = {\rm const},\, \pi
(f) = a\, {\rm id}_H \]
is satisfied trivially.

This means that if $(G,U,E_{m_1}, \pi )$ is the local S.I.
at the malitiously singular point $pt$, then the condition for S.I.
does not impose any limitations on the representation $(U,H)$. In
particular, it can be an irreducible representation. $\Box $

\section{Example: Two-Dimensional RWFL World Model}
In this section, we consider a simplified (two-dimensional) RWFL
cosmological model with its two malicious singularities that
often serves as a typical example in the classical singularity
problem (Bosshard, 1976; Dodson, 1978).

Let us consider the spacetime
\[M=\{(\eta ,\chi ):\eta\in (0,T),\chi\in 
S^1\},\] 
where $(0,T)\subset {\bf R}$, carrying the metric
\[ds^2=R^2(\eta )(-d\eta^2+d\chi^2).\]
This model has the initial singularity: $R^2(\eta )\rightarrow
0$ as $\eta\rightarrow 0$, and the final singularity: $R^2(\eta
)\rightarrow 0$ as $\eta\rightarrow T$ (for the detailed
presentation of this model see (Dodson 1978) or (Heller and
Sasin 2002).
\par
To make a contact with our previous construction, let us list
all relevant magnitudes:

$M =(0,T)\times S^1$,

$(\eta, \chi , \lambda ) \in E$, $t \in {\bf R}$,

$\gamma = (\eta, \chi , \lambda, t) \in \Gamma $,

$d(\gamma ) = (\eta, \chi , \lambda )$,

$r(\gamma ) = (\eta, \chi , \lambda + t)$,

$\Gamma^p = \{ (pt^{-1},t): t \in {\cal R} \} = \{(\eta, \chi ,
\lambda - t, t \}$.

To obtain the groupoid representation corresponding to a given
representation $(U,H)$ of the group $G \simeq {\cal R}$, we
construct the Hilbert space for a chosen regular point $p_0 =
(\eta , \chi, t_0)$

\[{\cal H} = \{F: \Gamma^{p_0} \rightarrow H: F(p_0g^{-1},g) =
U(g)F(p_0,e)\} \]
\[ = \{ F(\eta, \chi, \lambda_0 - t,t) = U(t)F(\eta , \chi ,
\lambda_0 , 0)\}.\]

And for the groupoid representation operator we have 
\[{\cal U}(p_0,g_0)F:= {\cal U}(\eta , \chi , \lambda_0 , t_0
)F(\eta , \chi , \lambda_0 + t_0 - t, t) \]
\[ = F(\eta, \chi , \lambda_0+t_0 - t, t- t_0) = U(-t_0)F(\eta ,
\chi , \lambda_0 - t,t).\]

To obtain the corresponding S.I. $(G,U,E_{(\eta , \chi )}, P)$,
for $G = {\bf R}$, we make use of the generalized Stone theorem
[Neimark, Ambrose, Godement; see Barut and R\c{a}czka, 1977,
p. 160)] which says that a representation $(U,H)$ of the group
${\bf R}$ in any Hilbert space can be expressed, with the help
of a spectral measure $P$, in the following way 
\[ U(t) = \int_{\bf R}e^{its}dP(s).\]

We have
\[{\cal F}_G(E_m,H) = \{\psi : E_m \rightarrow H: \psi (\eta ,
\chi , \lambda_0 + t) = U(t)\psi (\eta ,\chi , \lambda_0)\} .\]
In this Hilbert space the spectral measure is
\[ P(B)\psi(\eta , \chi , t) = \chi_B(t) \psi(\eta , \chi , t)\]
where $B \subset {\bf R}$ is a Borel set, and $\chi_B$ its
characteristic function.

It can be easily seen that the system $(G,U, E_m, P)$ indeed
satisfies conditions of S.I. Therefore, the results obtained in
the previous sections remain valid. For regular points, the
representation $(U, H)$ is equivalent to the regular
representation of the group ${\bf R}$ in $L^2({\bf R})$,
possibly with the multiplicity greater than 1. For malicious
singularities, every representation $(U,H)$ of the group ${\bf
R}$ satisfies the conditions of S.I. The regular representation
of ${\bf R}$ in $L^2({\bf R})$ has, exactly as for $SO(3,1)$, no
discrete components.

\section{Interpretation and Comments}
So far  our results were purely formal; let us now try to read
from them a physical meaning. In physical applications systems
of imprimitivity appear in the following circumstances.

Let us consider a quantum physical system having the symmetry
group $P$. It is described by a pair $(U(P),H)$ where $U(P)$ is
a unitary representation of the group $P$ in a Hilbert space
$H$.  Let us further assume that a classical system is described
by the pair $(P,M)$, where $M$ is the space of a classical
observable that charctersizes the state of this system (e.g. the
space of positions or space of momenta), and $P$ is acting
on $M$ as its symmetry group. If in $H$ there is a state
$\psi_x$ in which the value of an observable is $x\in M$, we say
that the quantum state $\psi_x$ corresponds to the classical
magnitude $x$. Let us denote
\[ H_x = \{\psi_a \in H: a = x \}. \]
If such correspondence exists, i.e., if the quantum system has
an interpretation in terms of classical observables, the
following conditions hold

(i) $H_x = \bigcup_{x \in M}H_x,$

(ii) $U(p)H_x \subset H_{px}$,

\noindent
and there exists the system of imprimitivity for the
representation $U(p)$ of the symmetry group $P$ (Mensky, 1976a,
b). The above is visualized in the following diagram, the left
column of which represents quantum description and its right
column the corresponding classical description.

\unitlength=1mm
\special{em:linewidth 0.4pt}
\linethickness{0.4pt}
\begin{picture}(105.00,110.00)
\put(35.00,110.00){\makebox(0,0)[rc]{$H_x \ni \psi_x$}}
\put(65.00,110.00){\vector(-1,0){25.00}}
\put(65.00,110.00){\vector(1,0){30.00}}
\put(100.00,110.00){\makebox(0,0)[lc]{$x\in M$}}
\put(35.00,105.00){\vector(0,-1){30.00}}
\put(35.00,70.00){\makebox(0,0)[rc]{$U(p)\psi_x$}}
\put(100.00,105.00){\vector(0,-1){30.00}}
\put(100.00,70.00){\makebox(0,0)[lc]{$px\in M$}}
\put(65.00,70.00){\vector(-1,0){25.00}}
\put(65.00,70.00){\vector(1,0){30.00}}
\put(30.00,90.00){\makebox(0,0)[rc]{$p\in P$}}
\put(105.00,90.00){\makebox(0,0)[lc]{$p \in P$}}
\end{picture}
\vspace{-6cm}

If $P$ acts on $M$ transitively, i.e., if there is a subgroup
$K\subset P$ such that $M = K\backslash P$, then, on the
strength of the Mackey theorem, any imprimitive representation
of the group $P$ is induced from the subgroup $K$ (Mackey, 1978,
1998).

Let us now apply this analysis to the case of spacetime with
malicious singularities. The groupoid representation is given by
the pair $({\cal U}, \{ H_u \}_{u\in E})$. Although in the
present work we consider classical singularities, we can say
that the above pair provides a quantum description of the
singularity (or something analogous to quantum description since
it uses typically quantum mathematical tools).  We also have its
classical description given by the action of the group $G =
SO(3,1)$ on $E$, $E \times G \rightarrow E$ (Lorentz rotations
of local frames). Since the groupoid representation ${\cal U}$
corresponds bijectively to the system of imprimitivity $(G,U,E,
\pi )$, we could say that the quantum description of our model
corresponds to its classical description. This somehow justifies
the fact that although we are facing the classical singularity
problem, it can be dealt with in terms of mathematical
structures typical for quantum theory (unitary operators,
Hilbert spaces, etc.).

In our case, the Mackey theorem says that the unitary
representation of the Lorentz group $G=SO(3,1)$, which is the
part of the corresponding S.I., is induced from its subgroup $K$
such that $E_m = K\backslash G$. If $m \in M$ is a regular point
then $K = \{ e \} $; if $m \in \bar M \setminus M$ is a
malicious singularity then $K = G$. This means that in our model
the correspondence between quantum description and classical
description is complete, if we do not take into account
malicious singularities. At maliciously singular points this
correspondence formally also takes place, but the S.I. condition
is always trivially satisfied.

There can exist ``intermediate'' singularities for which the
isotropy group $K$ is a proper subgroup of $G$; they are not
regular points of spacetime, but as singularities are weaker than malicious
ones [for examples see (Ellis and Schmidt, 1977)]. Let
$K_p$ be the isotropy group of a point $p\in E$. We have, $K_p =
K_q$ if there is $g \in G$ such that $q=pg$, and $E_m =
K_p\backslash G$. Since $E_m$, for an ``intermdiate''
singularity at $m$, is a quotient space, the Mackey theorem
applies, and consequently the represenation, that is a part of
the S.I. with the base $E_m$, is an induced representation by a
certain representation of the subgroup $K$.

Let us notice that if $m$ is a regular point of spacetime,
dim$E_m = {\rm dim}G$; if $m$ is an ``intermediate''
singularity, dim$E_m = $ dim$G - $dim$K$; if $m$ is a malicious
singularity, dim$E_m = 0$. In this sense, $K$ may be regarded as
measuruing the ``strength'' of a given singulatrity.

At regular points the group representation, which is an element
of S.I., does not have discrete components (the group $SO(3,1)$
has no discrete series). In the quantum field theory this
implies the impossibility to localize an elementary particle. At
the malicious singularity such a group representation can be a
single irreducible representation or a direct sum of such
representations. Formally speaking, this would mean that at the
singularity elementary particles can be localized. Since,
however this follows from the fact that the S.I. condition does
not impose any limitations on what can happen here, the correct
interpretation seems to be that general relativity is
essentially an incomplete theory: malicious singularities are
its ``open windows'' that claim for a more general (and more
complete) theory. This is not true, however, that we know
nothing about the nature of the malicious singularity; as we
have shown, some of its characteristics surrender to the
analysis in terms of representations in Hilbert spaces.

\vspace{1.5cm}

\noindent
{\bf REFERENCES\/}

\vspace{1.2cm}

\pn
Barut, A. O. and R\c{a}czka, R. (1980). {\it Theory of Group
Representations and Applications\/}, Polish Scientific
Publishers, Warsaw.
\pn
Bosshard, B. (1976). {\it Communications in Mathematical
Physics\/}m {\bf 46}, 263.
\pn
Connes, A. (1994). {\it Noncommutative Geometry\/}, Academic
Press, New York.
\pn
Dodson, C. T. J. (1978). {\it International Journal of
Theoretical Physics\/} {\bf 18}, 898.
\pn
Ellis, G. F. R. and Schmidt, B. G. (1977). {\it General
Relativity and Gravitation\/} {\bf 11}, 915.
\pn
Eardley, D. M. and Smarr, L. (1979). {\it Physical Review\/} {\bf
D19}, 2239.
\pn
Heller, M. and Sasin, W. (1994). {\it General Relativity and
Gravitation\/} {\bf 26}, 797.
\pn
Heller, M. and Sasin, W. (1996). {\it Journal of Mathematical
Physics\/} {\bf 37}, 5665.
\pn
Heller, M. and Sasin, W. (1999). {\it General Relativity and
Gravitation\/} {\bf 31}, 555.
\pn
Heller, M. and Sasin, W. (2002). {\it International Journal of
Theoretical Physics\/} {\bf 41}, 919.
\pn
Johnson, R. A. (1977). {\it Journal of Mathematical Physics}
{\bf 18}, 898. 
\pn
Landsman N. P, {\it Mathematical Topics between Classical and
Quantum Mechanics\/}, Springer, New York, 1998.
\pn
Mackey, G. W. (1952). {\it Annals of Mathematics\/} {\bf 55},
101. 
\pn
Mackey, G. W. (1978). {\it Unitary Group Representations in
Physics, Probability and Number Theory\/}, Benjamin, Counings.
\pn
Mackey, G. W. (1998). {\it Contemporary Mathematics\/} {\bf
214}, 91.
\pn
Mensky, M. B. (1976a). {\it Communications in Mathematical
Physics\/} {\bf 47}, 97.
\pn
Mensky, M. B. (1976b). {\it The Method of Induced Variables:
Spacetime and the Concept of Particle\/}, Nauka, Moscow (in
Russian). 
\pn
Paterson, A. L. T. (1999). {\it Groupoids, Inverse Semigroups,
and Their Operator Algebras\/}, Birkh\"auser, Boston - Basel -
Berlin.
\pn
Rudnicki, W., Budzy\'nski, R. J. and Kondracki, W. (2002) {\it
Modern Physics Letters\/} {\bf A17}, 387.
\pn
Schmidt, B. G. (1971). {\it General Relativity and
Gravitation\/} {\bf 1}, 269.
\pn
Tipler, F. J. (1977a). {\it Physics Letters\/} {\bf A64}, 8.
\pn
Tipler, F. J. (1977b). {\it Nature\/} {\bf 270}, 500.

\end{document}